\documentclass[%
 reprint,
superscriptaddress,
 amsmath,
 amssymb,
 aps,
 prl,
]{revtex4-1}

\usepackage{graphicx}% Include figure files
\usepackage{dcolumn}% Align table columns on decimal point
\usepackage{bm}% bold math
\usepackage{color}
\begin{document}

\preprint{APS/123-QED}

\title{Speeding up  dynamics by tuning the non-commensurate size of rod-like particles in a smectic phase}% Force line breaks with \\

\author{Massimiliano Chiappini}
\email{m.chiappini@uu.nl}
\affiliation{Soft Condensed Matter, Debye Institute for Nanomaterials Science, Department of Physics, Utrecht University, Princetonplein 1, Utrecht 3584 CC, The Netherlands}
\author{Eric Grelet}
\affiliation{Centre de Recherche Paul-Pascal, UMR 5031, CNRS \& Universit\'e de Bordeaux, 115 Avenue Schweitzer, 33600 Pessac, France}
\author{Marjolein Dijkstra}
\email{m.dijkstra@uu.nl}
\affiliation{Soft Condensed Matter, Debye Institute for Nanomaterials Science, Department of Physics, Utrecht University, Princetonplein 1, Utrecht 3584 CC, The Netherlands}

\date{\today}% It is always \today, today,
\begin{abstract}
Using simulations, we study the diffusion of rod-like guest particles in a smectic environment of rod-like host particles. We find that the dynamics of  guest rods across smectic layers changes from a fast nematic-like diffusion to a slow hopping-type dynamics via an intermediate switching regime  by varying the length of the guest rods with respect to the smectic layer spacing. We determine the optimal rod length that yields the fastest and the slowest diffusion in a lamellar environment. We show that this behavior can be rationalized by a complex 1D effective periodic potential exhibiting two energy barriers, resulting in a varying preferred mean position of the guest particle in the smectic layer. The interplay of these two barriers controls the dynamics of the guest particles yielding a slow, an intermediate and a fast diffusion regime depending on the particle length.
\end{abstract}

\pacs{Valid PACS appear here}% PACS, the Physics and Astronomy
\maketitle

Understanding the dynamics of particles or objects in crowded environments is important in many fields ranging from traffic jams \cite{Nag2002},  evacuations of crowds, sheep herding, evasive tumor growth, to caging in colloidal glasses \cite{Wee2002,Poo2004,Cha2007}. The  motion of a guest particle in a disordered crowded environment is severely hampered by its surrounding constituents. As most disordered systems are characterized by only one relevant length scale (\emph{e.g.} particle size), a simple picture emerges: the bigger the particle the slower its dynamics \cite{Smi1999,Gol2006,Dix2008,Sok2012}. This phenomenon is invariant across scales as demonstrated by the above-mentioned examples. However, this simple picture breaks down as the environment becomes inhomogeneous and ordered, yielding additional competing length scales and giving rise  to remarkable exceptions to this general rule. 

The motion of particles in ordered environments has been thoroughly studied in the field of liquid crystals, finding that crowded environments with different degrees of positional and/or orientational order lead to a wide variety of dynamic behaviors. For nematic liquid crystals, exhibiting long-range orientational order, the anisotropy of the environment is transferred to the motion of the particles. 
\begin{figure*}[hbtp]
\centering
\includegraphics[width=\textwidth]{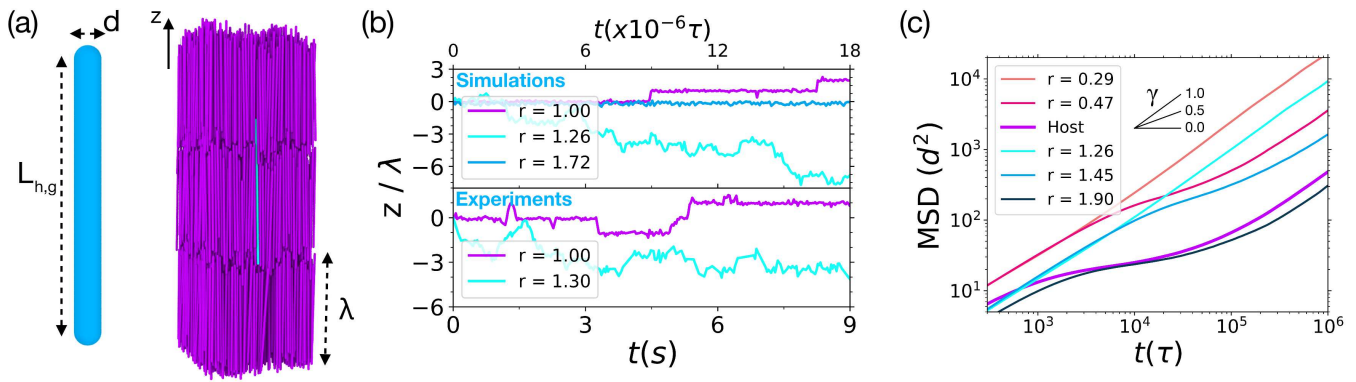}
\caption{(a) Snapshot from simulations of a guest spherocylinder (cyan) with cylindrical length $L_{g}$ and diameter $d$ diffusing in a host smectic phase of layer spacing $\lambda$ formed by hard spherocylinders (purple) with equal diameter $d$ and length $L_{h} = 40 d$. (b) Example trajectories of guest particles with varying size ratio $r = (L_{g} + d) / \lambda$ along the nematic director $\hat{\mathbf{n}}$ of the host smectic phase in simulations (top) and experiments (bottom) \cite{Alv2017} showing the fast nematic-like diffusion of non-commensurate guest rods with $r \sim 1.3$ and discrete hopping-type diffusion of host particles ($r \sim 1$). The conversion factor from the computational time unit $\tau$ to seconds $s$ ($\tau \sim 2 \cdot 10^{-6} s$) is discussed in the Supplemental Material \cite{SM}. (c) Longitudinal mean square displacement (MSD) of simulated guest particles of varying size ratios $r = (L_{g} + d) / \lambda$, showing either a fast nematic-like diffusion for non-commensurate guest rods of $r \sim 1.3$ and $0.3$, or a sub-diffusive regime for the other guest and host particles. The diffusion exponents  $\gamma= 0.5$ and $1$ are indicated for comparison (See Eq. (1)).
	 \label{fig:Figure01}}
\end{figure*}
A fast longitudinal self-diffusion is observed in the direction parallel to the nematic director ${\bf \hat{n}}$ (the average particle orientation), and a slow transverse self-diffusion in the perpendicular direction \cite{Bru1998,Low1999,Let2005}. 

In the case of long-range positional order, the dynamics strongly depends on the dimensionality of the translational order and the corresponding effective energy landscape. In 3D colloidal crystals, particles are confined to their lattice positions, and the diffusion is largely determined by the motion of defects \cite{Als2005,Meer2014,Mee2017}. In columnar liquid crystals, showing 2D positional order, a liquid-like longitudinal diffusion is observed within the columns, accompanied by a transverse hopping-type dynamics between different columns \cite{Bel2010,Nad2013}. Finally, in smectic liquid crystal phases characterized by a quasi long-range 1D translational order, a quantized hopping-type dynamics is found across smectic layers as the particles experience an effective one-dimensional periodic potential due to the lamellar organization \cite{Let2007,Mat2010,Pou2011}. Furthermore, computer simulations demonstrated cooperative motion of string-like clusters of particles across the smectic layers \cite{Pat2010}.

In general, the presence of positional and/or orientational order introduces additional length scales to the system. In the presence of guest particles, their interplay with the various length scales associated with the structure increases the complexity of the dynamics. On the one hand, the diffusion of guest spherical particles in nematic phases of rod-like host liquid crystals has been widely addressed in literature  \cite{Ruh1996,Sta2001,Lou2004,Kan2005,Mon2012,Tur2013,Mar2018}, finding a faster diffusion in the direction longitudinal to the nematic director field. On the other hand, the diffusion of non-spherical particles in anisotropic liquid crystalline environments is still largely unexplored. Recently, Alvarez \emph{et al.} \cite{Alv2017} studied in experiments the diffusion of tracer amounts of non-commensurate guest viral rods in a smectic phase of shorter host fd filamentous viruses with a size ratio $L_\text{guest}/L_\text{host} \simeq 1.3$. Surprisingly they found that while the host particles experience the usual hopping-type dynamics across smectic layers, the non-commensurate guest particles undergo a fast and almost continuous nematic-like diffusion, yielding the exceptional case of larger guest particles diffusing faster than the smaller host ones. No significant differences between host and guest particles were found in the transverse in-layer diffusion. The typical slow hopping-type diffusion across smectic layers was recovered for dimeric and trimeric mutants of the host fd particles, namely for guest particles with length ratios of 2 and 3, respectively.

In this Letter, we study using computer simulations the dynamics of guest particles of varying lengths in a smectic environment of host particles in order to unravel the mechanism behind this highly counterintuitive fast diffusion of large non-commensurate guest particles. We show that  by tuning the length of the guest rods with respect to the smectic layer spacing their longitudinal dynamics changes from a fast nematic-like diffusion to a slow hopping-type dynamics via an intermediate switching regime, thereby obtaining control over the speed and type of behavior of the longitudinal diffusion. More importantly, we determine the optimal rod size for either the fastest or slowest diffusion, and rationalize this behavior in terms of a complex 1D effective smectic periodic potential characterized by two energy barriers that each rod feels in the lamellar structure of the smectic phase. We show that the interplay and relative height of the two energy barriers control the dynamics of the guest particles, yielding a slow, an intermediate and a fast diffusion regime depending on the particle length. 

We model the experimental mixture of long and short filamentous bacteriophage viruses as a binary mixture of rigid rods. Each guest and host rod is modeled by a hard spherocylinder, i.e. a cylinder of diameter $d$ and length $L_g$ and $L_h$, respectively, capped at both ends with hemispheres of diameter $d$, yielding an end-to-end length of $L_{g,h} + d$ (Fig. \ref{fig:Figure01}a). We introduce a tracer amount of $N_g = 6$  guest particles in a system of $N_h = 3072$ host particles with a length $L_h = 40d$. The overall phase sequence  of isotropic, nematic, smectic-A (Sm$_A$), and smectic-B/crystal phases of fd-viruses \cite{Grel2014} is well captured by that  of hard spherocylinders with $L_h=40d$ \cite{Bol1997}, even though fd virus suspensions also display a columnar phase \cite{Gre2008}. The aspect ratio of the host rods in the simulations is set such that it roughly matches the effective rod length over diameter ratio of the experimental system, thereby taking into account the electrostatic repulsion of the fd viruses \cite{Grel2014}. 

We equilibrate the system in a low-density Sm$_A$ state  using Monte Carlo (MC) simulations in an isothermal-isobaric ensemble, i.e. the pressure, temperature, $N_g$ and $N_h$ are kept fixed. Note that the smectic layer spacing in simulations is $\lambda \sim 1.1 L_{h}$, whereas $\lambda \sim 1.0 L_\text{host}$ in the experimental system of filamentous viruses \cite{Grel2014}. After full equilibration we investigate using both standard and Dynamic MC (DMC) simulations \cite{Pat2012,Pat2015} the longitudinal dynamics along the $z$-axis, parallel to the nematic director $\hat{\mathbf{n}}$, for various $L_{g} \in [0.2,2.5] L_h$ corresponding to various size ratios $r = (L_{g} + d) / \lambda$. Within this range of lengths the probability of finding guest rods in a transverse inter-lamellar configuration is negligible \cite{Roi1995, Dui1997}. We refer the reader to the Supplemental Material~\cite{SM} for technical details on the simulations. 

In Fig. \ref{fig:Figure01}b, we present typical longitudinal trajectories from both simulations and experiments, showing remarkably similar slow hopping-type dynamics of host particles ($r\sim 1$) as well as fast diffusive behavior of non-commensurate guest particles ($r\sim 1.3$). For each particle trajectory $z(t)$ we measure the mean square displacement along the director $\hat{\mathbf{n}}$, MSD$(t) = \langle (z(t_0+t) - z(t_0))^2\rangle$, and average the MSDs of all particles with equal length. In Fig.~\ref{fig:Figure01}c we show the MSDs for a selected set of size ratios $r$. For particles with a length commensurate with the smectic layer spacing ($r \sim 1$) we obtain the typical MSD of particles in a lamellar phase \cite{Mat2010} with a cage-trapping plateau between the short- and long-time diffusion regimes corresponding to the intra- and inter-layer dynamics, respectively (See Supplemental Material~\cite{SM}). As the length of the guest particles increases, the time interval for the caging becomes shorter, and eventually disappears for $r \sim 1.25$ when the dynamics becomes nematic-like with a diffusive behavior  (Fig.~\ref{fig:Figure01}c). Upon further increasing the particle length, the cage-trapping plateau reappears ($r\sim 1.45$) and becomes more pronounced as the dynamics becomes hopping-type again for nearly commensurate dimers ($r\sim 1.90$). Similarly, for guest rods shorter than the smectic layer spacing, the time interval of caging decreases ($r\sim 0.47$) and eventually disappears for guest particles of low anisotropy ($r \sim 0.29$).

\begin{figure}[h!tpb]
\centering
\includegraphics[width=0.5\textwidth]{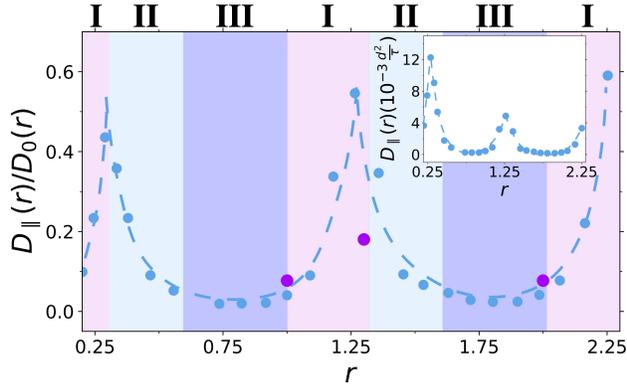}
\caption{Long-time diffusion coefficient $D_{\parallel}(r)$ normalized by the infinite-dilution diffusion coefficient $D_0(r)$ \cite{SM} of guest particles as a function of the size ratio $r$. The experimental values of $D_{\parallel}(r)$ are shown in purple for $r = 1$, $1.3$, and $2$ \cite{Alv2017}. The inset shows the raw diffusion coefficients. The background is colored according to the three diffusion regimes displayed in Fig. \ref{fig:Figure03}, and the dashed lines are guides to the eye.
\label{fig:Figure02}}
\end{figure}

To quantify the long-term dynamic behavior, we determine the long-time diffusion coefficient $D_{\parallel}$ defined as half the slope of the MSD at long times, \emph{i.e.} MSD$(t) = 2D_{\parallel} t^{\gamma}$ (1), and we present $D_{\parallel}$ normalized by the particle diffusion coefficient at infinite dilution $D_0(r)$ as a function of the size ratio $r$ in  Fig.~\ref{fig:Figure02}. In the range $1 \leq r < 2$, a strong increase of the diffusion is observed with a maximum $D_{\parallel}(r)/D_0(r)$ at $r \sim 1.25$, corresponding to a fast nematic-like diffusion of particles whose length is not commensurate with the smectic layer spacing. This yields an optimal value for the fastest longitudinal diffusion remarkably close to the particle length ratio for which fast diffusion was observed in experiments~\cite{Alv2017}. For larger $r$ the diffusion slows down as the hopping-like dynamics is retrieved. The slowest diffusion is not found for particles twice the length of the smectic layer spacing ($r \sim 2$) but at slightly smaller lengths ($r \sim 1.75$). We also observe in Fig.~\ref{fig:Figure02} that the values of  $D_{\parallel}(r)/D_0(r)$ are in good quantitative agreement with the experimental values marked by the purple symbols despite the simplicity of our model. For guest particles shorter than the host ones ($r < 1$), the fastest and the slowest dynamics are obtained by non-commensurate particles of size ratio $r \sim 0.25$ and $r \sim 0.75$ respectively, corresponding to the fast nematic-like diffusion for the former and slow hopping-like dynamics for the latter. Interestingly, the normalized values for  $r < 1$  of the diffusion coefficients for the slowest and fastest dynamics are very similar to their corresponding values for $r > 1$, emphasizing again that the shortest particles do not necessarily diffuse the fastest. In the long rod limit, i.e. for $r>2$, we find another maximum of $D_{\parallel}(r)/D_0(r)$ at $r \sim 2.25$. 

\begin{figure*}[h!tpb]
\centering
\includegraphics[width=\textwidth]{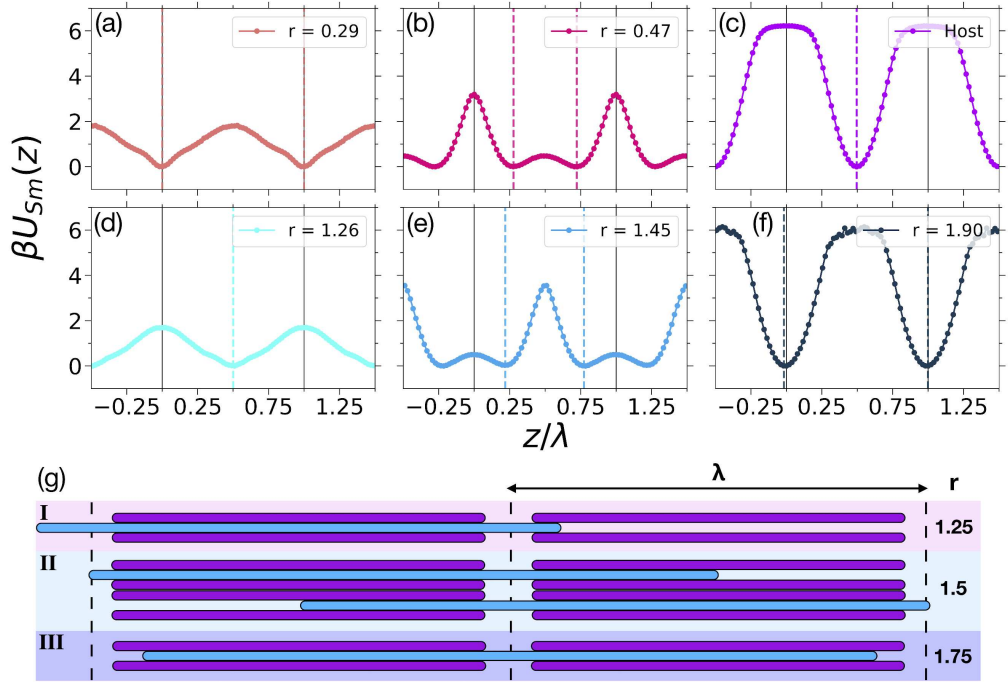}
\caption{(a-f) Effective potential $U_{Sm}(z)$ experienced by guest particles for varying size ratios $r = (L_g+d)/\lambda$ in a smectic phase with a layer spacing $\lambda$. The dashed vertical lines indicate the equilibrium positions of the rod particles, $z^{\textrm{\tiny min}}_1$ and $z^{\textrm{\tiny min}}_2$, corresponding to the minima of the ordering potential $U_{Sm}(z)$. A video showing the variation of $U_{Sm}(z)$ with the size ratio $r$ can be found in the SM \cite{SM}. (g) Sketches of the host (purple) and guest (cyan) particles at their equilibrium positions $z^{\textrm{\tiny min}}_1$ and $z^{\textrm{\tiny min}}_2$ for three exemplary size ratios ($r = 1.25, 1.5, 1.75$) corresponding to the different diffusion regimes.
\label{fig:Figure03}}
\end{figure*}

The dependence of $D_{\parallel}(r)/D_0(r)$ on the size ratio $r$  in Fig.~\ref{fig:Figure02} suggests a periodic behavior of the longitudinal dynamics with a period set by the smectic layer spacing $\lambda$. For each size ratio interval $r \in \lbrack n, n+1 \rbrack$ with $n =0,1,2, \cdots$, the dynamics first speeds up as $r$ increases and the smectic caging becomes less severe,  reaches a maximum value at  $r \simeq n+0.25$ corresponding to the fastest nematic-like diffusion, and then slows down and  reaches a minimal value at $r  \simeq n+0.75$. This periodic behavior can be explained by dividing the end-to-end guest rod length $L_g+d=r \lambda$  into a  length $\ell\lfloor r \rfloor $ that is commensurate with $\lfloor r \rfloor$ smectic layers (where the floor function $\lfloor x \rfloor$ denotes the largest integer that is less than $x$), and an ``excess"  length of $\ell(r - \lfloor r \rfloor) $. The longitudinal dynamics of guest particles is predominately determined by the excess part of the guest rod, which creates voids in the smectic layers and affects the caging of the lamellar phase. Here, the only effect of the ``commensurate" part of the particle is a general slowing down of the dynamics with $n$ (See the inset of Fig.~\ref{fig:Figure02}).

\begin{figure}[b!htp]
\centering
\includegraphics[width=0.5\textwidth]{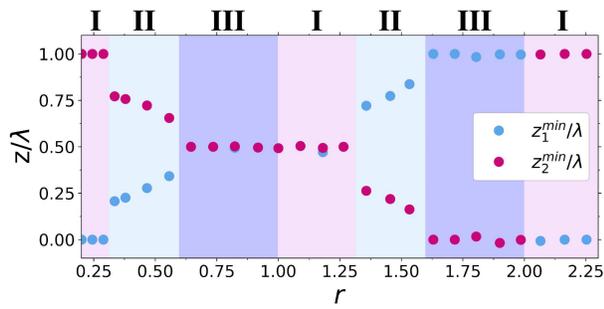}
\caption{ Center-of-mass positions  $z^{\textrm{\tiny min}}_{1}$ and $z^{\textrm{\tiny min}}_{2}$ of the guest rods corresponding to the minima of the effective smectic potential as a function of  size ratio $r$. The background is colored according to the three diffusive regimes displayed in Fig. \ref{fig:Figure03}. 
\label{fig:Figure04}}
\end{figure}

To quantify the effect of the excess particle length, we measure the effective potential $\beta U_{Sm}(z) = -\ln(\rho(z))$ felt by a guest rod, where $\rho(z)$ is the probability distribution of finding a rod-shaped particle in an infinitesimal interval of  $[z, z+\delta z]$ and $\beta = 1/k_BT$. The effective potential is periodic due to the smectic host ordering, therefore $\rho(z)$ is only measured in a single smectic layer $0 \leq z < \lambda$. In Fig.~\ref{fig:Figure03}a-f, we report the smectic potential for varying length ratios $0 < r < 2$. Surprisingly, we find that the smectic potentials exhibit two barriers, or equivalently two minima at $z^{\textrm{\tiny min}}_{1}$ and $z^{\textrm{\tiny min}}_{2}$ which merge into a single minimum when $r \simeq n$, namely when particles are commensurate with the layer spacing. We plot $z^{\textrm{\tiny min}}_{1}$ and $z^{\textrm{\tiny min}}_{2}$ for varying  $r$ in Fig.~\ref{fig:Figure04}, allowing us to distinguish three different regimes as schematically illustrated in Fig.~\ref{fig:Figure03}g. 

In the first regime ($\text{I}$) corresponding to size ratios $r\in \lbrack n, n+0.3 \rbrack$, the guest particles are on average located at the same position as their commensurate counterparts with $r = n$, i.e. in the middle of the smectic layers. However, as they are longer than $n\lambda$, they create holes in the adjacent smectic layers resulting in a release of the cage constraint thereby facilitating the inter-layer diffusion and speeding up the dynamics, with the fastest nematic-like diffusion found for $r \sim n + 0.25$. In the opposite limit, the third regime ($\text{III}$) having $r \in \lbrack n+0.6, n+1 \rbrack$ exhibits the slowest diffusion behavior and corresponds to guest particles which already have the same equilibrium position as the next commensurate multimer ($r = n + 1$). Because the guest rods are shorter than $(n + 1)\lambda$, they first have to diffuse within the smectic layer to reach one of its two boundaries, before they can jump to the adjacent layer, slowing down the longitudinal diffusion in comparison to the one associated with commensurate particles. 
We denote regime $\text{III}$ as the slow diffusive regime. More intriguingly perhaps is the regime $\text{II}$ with $r \in \lbrack n+0.3, n+0.6 \rbrack$, where the minima  $z^{\textrm{\tiny min}}_{1}$ and $z^{\textrm{\tiny min}}_{2}$ correspond to the center-of-mass positions at which one of the ends of the guest particles touches one of the boundaries of the smectic layers (Fig.~\ref{fig:Figure03}b and e). This was recently experimentally observed for short rods dispersed in colloidal monolayer of host rod-shaped particles with a length ratio $r \sim 0.5$ \cite{Sia2019}: the short rods were found to strongly anchor to the membrane interfaces, and only occasionally hop to the opposite interface. Our results confirm this anchoring behavior and extend it to particles even larger than the lamellar spacing. The preferential adsorption of non-commensurate guest rods at the interface of smectic layers can be explained by the fact that guest rods at the interface generates large voids that can be partially filled via small angular fluctuations of neighboring host particles, thereby hindering their diffusion. However, if the guest particle is at the center of a smectic layer (or in between two smectic layers), the resulting voids are smaller, making it harder for host particles to occupy the empty space. This would indeed require higher tilt angle of the host rods, hence generating a defect structure in the smectic organization. As a consequence, the guest particles escape from this central position and adhere to one of the two smectic layer interfaces. In this regime $\text{II}$, referred as the switching regime, the guest particles experience two potential barriers of varying height (Fig.~\ref{fig:Figure03}b and d) for varying $r$, which results from a non-trivial interplay of the effective smectic potentials that are felt  by  single host rods ($r\sim 1$, Fig.~\ref{fig:Figure03}c) as well as by commensurate rods ($r\sim 2$, Fig.~\ref{fig:Figure03}f) and which are out-of-phase in terms of barrier locations (See Supplemental Material~\cite{SM}).

In conclusion, we showed that the dynamics of guest rods can be controlled by tuning the ratio $r$ of their size over  %$r=(L_g + d)/\lambda$, with $\lambda$ 
the lamellar spacing. We observed that the long-time diffusion coefficient $D_{\parallel}$ is a periodic function of $r$, as the longitudinal dynamics is entirely determined by the excess length $\ell(r - \lfloor r \rfloor) $ of the guest particle. We show that this behavior can be rationalized by a 1D effective periodic potential exhibiting up to two energy barriers, yielding a slow, an intermediate and a fast diffusive regime,
% based on their length. In particular, the interplay and relative height of the two potential energy barriers control the dynamics of the guest particles, which changes from a slow hopping-type dynamics for a length ratio of $r \simeq n + 0.75$ to a fast nematic-like diffusion for $r \simeq n + 0.25$
granting complete control over the type and speed of the %longitudinal
dynamics of guest particles in a smectic environment.

\nocite{Fre2001}
\nocite{Tir1979}

\begin{acknowledgments}
M. C. and M. D. acknowledge financial support from the EU H2020-MSCA-ITN- 2015
project ‘MULTIMAT’ (Marie Sklodowska-Curie Innovative
Training Networks) [project number: 676045]. 
\end{acknowledgments}

\bibliography{Bibliography}

\begin{thebibliography}{39}%
\makeatletter
\providecommand \@ifxundefined [1]{%
 \@ifx{#1\undefined}
}%
\providecommand \@ifnum [1]{%
 \ifnum #1\expandafter \@firstoftwo
 \else \expandafter \@secondoftwo
 \fi
}%
\providecommand \@ifx [1]{%
 \ifx #1\expandafter \@firstoftwo
 \else \expandafter \@secondoftwo
 \fi
}%
\providecommand \natexlab [1]{#1}%
\providecommand \enquote  [1]{``#1''}%
\providecommand \bibnamefont  [1]{#1}%
\providecommand \bibfnamefont [1]{#1}%
\providecommand \citenamefont [1]{#1}%
\providecommand \href@noop [0]{\@secondoftwo}%
\providecommand \href [0]{\begingroup \@sanitize@url \@href}%
\providecommand \@href[1]{\@@startlink{#1}\@@href}%
\providecommand \@@href[1]{\endgroup#1\@@endlink}%
\providecommand \@sanitize@url [0]{\catcode `\\12\catcode `\$12\catcode
  `\&12\catcode `\#12\catcode `\^12\catcode `\_12\catcode `\%12\relax}%
\providecommand \@@startlink[1]{}%
\providecommand \@@endlink[0]{}%
\providecommand \url  [0]{\begingroup\@sanitize@url \@url }%
\providecommand \@url [1]{\endgroup\@href {#1}{\urlprefix }}%
\providecommand \urlprefix  [0]{URL }%
\providecommand \Eprint [0]{\href }%
\providecommand \doibase [0]{http://dx.doi.org/}%
\providecommand \selectlanguage [0]{\@gobble}%
\providecommand \bibinfo  [0]{\@secondoftwo}%
\providecommand \bibfield  [0]{\@secondoftwo}%
\providecommand \translation [1]{[#1]}%
\providecommand \BibitemOpen [0]{}%
\providecommand \bibitemStop [0]{}%
\providecommand \bibitemNoStop [0]{.\EOS\space}%
\providecommand \EOS [0]{\spacefactor3000\relax}%
\providecommand \BibitemShut  [1]{\csname bibitem#1\endcsname}%
\let\auto@bib@innerbib\@empty
%</preamble>
\bibitem [{\citenamefont {Nagatani}(2002)}]{Nag2002}%
  \BibitemOpen
  \bibfield  {author} {\bibinfo {author} {\bibfnamefont {T.}~\bibnamefont
  {Nagatani}},\ }\href {\doibase 10.1088/0034-4885/65/9/203} {\bibfield
  {journal} {\bibinfo  {journal} {Reports on Progress in Physics}\ }\textbf
  {\bibinfo {volume} {65}},\ \bibinfo {pages} {1331} (\bibinfo {year}
  {2002})}\BibitemShut {NoStop}%
\bibitem [{\citenamefont {.Weeks}\ and\ \citenamefont {Weitz}(2002)}]{Wee2002}%
  \BibitemOpen
  \bibfield  {author} {\bibinfo {author} {\bibfnamefont {E.~R.}\ \bibnamefont
  {.Weeks}}\ and\ \bibinfo {author} {\bibfnamefont {D.~A.}\ \bibnamefont
  {Weitz}},\ }\href {\doibase https://doi.org/10.1016/S0301-0104(02)00667-5}
  {\bibfield  {journal} {\bibinfo  {journal} {Chemical Physics}\ }\textbf
  {\bibinfo {volume} {284}},\ \bibinfo {pages} {361 } (\bibinfo {year}
  {2002})},\ \bibinfo {note} {strange Kinetics}\BibitemShut {NoStop}%
\bibitem [{\citenamefont {Poon}(2004)}]{Poo2004}%
  \BibitemOpen
  \bibfield  {author} {\bibinfo {author} {\bibfnamefont {W.~C.~K.}\
  \bibnamefont {Poon}},\ }\href {\doibase 10.1557/mrs2004.35} {\bibfield
  {journal} {\bibinfo  {journal} {MRS Bulletin}\ }\textbf {\bibinfo {volume}
  {29}},\ \bibinfo {pages} {96–99} (\bibinfo {year} {2004})}\BibitemShut
  {NoStop}%
\bibitem [{\citenamefont {Chaudhuri}\ \emph {et~al.}(2007)\citenamefont
  {Chaudhuri}, \citenamefont {Berthier},\ and\ \citenamefont {Kob}}]{Cha2007}%
  \BibitemOpen
  \bibfield  {author} {\bibinfo {author} {\bibfnamefont {P.}~\bibnamefont
  {Chaudhuri}}, \bibinfo {author} {\bibfnamefont {L.}~\bibnamefont {Berthier}},
  \ and\ \bibinfo {author} {\bibfnamefont {W.}~\bibnamefont {Kob}},\ }\href
  {\doibase 10.1103/PhysRevLett.99.060604} {\bibfield  {journal} {\bibinfo
  {journal} {Phys. Rev. Lett.}\ }\textbf {\bibinfo {volume} {99}},\ \bibinfo
  {pages} {060604} (\bibinfo {year} {2007})}\BibitemShut {NoStop}%
\bibitem [{\citenamefont {Smith}\ \emph {et~al.}(1999)\citenamefont {Smith},
  \citenamefont {Morrison}, \citenamefont {Wilson}, \citenamefont
  {Fern{\'a}ndez},\ and\ \citenamefont {Cherry}}]{Smi1999}%
  \BibitemOpen
  \bibfield  {author} {\bibinfo {author} {\bibfnamefont {P.~R.}\ \bibnamefont
  {Smith}}, \bibinfo {author} {\bibfnamefont {I.~E.~G.}\ \bibnamefont
  {Morrison}}, \bibinfo {author} {\bibfnamefont {K.~M.}\ \bibnamefont
  {Wilson}}, \bibinfo {author} {\bibfnamefont {N.}~\bibnamefont
  {Fern{\'a}ndez}}, \ and\ \bibinfo {author} {\bibfnamefont {R.~J.}\
  \bibnamefont {Cherry}},\ }\href {\doibase
  https://doi.org/10.1016/S0006-3495(99)77486-2} {\bibfield  {journal}
  {\bibinfo  {journal} {Biophysical Journal}\ }\textbf {\bibinfo {volume}
  {76}},\ \bibinfo {pages} {3331 } (\bibinfo {year} {1999})}\BibitemShut
  {NoStop}%
\bibitem [{\citenamefont {Golding}\ and\ \citenamefont {Cox}(2006)}]{Gol2006}%
  \BibitemOpen
  \bibfield  {author} {\bibinfo {author} {\bibfnamefont {I.}~\bibnamefont
  {Golding}}\ and\ \bibinfo {author} {\bibfnamefont {E.~C.}\ \bibnamefont
  {Cox}},\ }\href {\doibase 10.1103/PhysRevLett.96.098102} {\bibfield
  {journal} {\bibinfo  {journal} {Phys. Rev. Lett.}\ }\textbf {\bibinfo
  {volume} {96}},\ \bibinfo {pages} {098102} (\bibinfo {year}
  {2006})}\BibitemShut {NoStop}%
\bibitem [{\citenamefont {Dix}\ and\ \citenamefont {Verkman}(2008)}]{Dix2008}%
  \BibitemOpen
  \bibfield  {author} {\bibinfo {author} {\bibfnamefont {J.~A.}\ \bibnamefont
  {Dix}}\ and\ \bibinfo {author} {\bibfnamefont {A.~S.}\ \bibnamefont
  {Verkman}},\ }\href {\doibase 10.1146/annurev.biophys.37.032807.125824}
  {\bibfield  {journal} {\bibinfo  {journal} {Annual Review of Biophysics}\
  }\textbf {\bibinfo {volume} {37}},\ \bibinfo {pages} {247} (\bibinfo {year}
  {2008})},\ \bibinfo {note} {pMID: 18573081}\BibitemShut {NoStop}%
\bibitem [{\citenamefont {Sokolov}(2012)}]{Sok2012}%
  \BibitemOpen
  \bibfield  {author} {\bibinfo {author} {\bibfnamefont {I.~M.}\ \bibnamefont
  {Sokolov}},\ }\href {\doibase 10.1039/C2SM25701G} {\bibfield  {journal}
  {\bibinfo  {journal} {Soft Matter}\ }\textbf {\bibinfo {volume} {8}},\
  \bibinfo {pages} {9043} (\bibinfo {year} {2012})}\BibitemShut {NoStop}%
\bibitem [{\citenamefont {Alvarez}\ \emph {et~al.}(2017)\citenamefont
  {Alvarez}, \citenamefont {Lettinga},\ and\ \citenamefont {Grelet}}]{Alv2017}%
  \BibitemOpen
  \bibfield  {author} {\bibinfo {author} {\bibfnamefont {L.}~\bibnamefont
  {Alvarez}}, \bibinfo {author} {\bibfnamefont {M.~P.}\ \bibnamefont
  {Lettinga}}, \ and\ \bibinfo {author} {\bibfnamefont {E.}~\bibnamefont
  {Grelet}},\ }\href {\doibase 10.1103/PhysRevLett.118.178002} {\bibfield
  {journal} {\bibinfo  {journal} {Phys. Rev. Lett.}\ }\textbf {\bibinfo
  {volume} {118}},\ \bibinfo {pages} {178002} (\bibinfo {year}
  {2017})}\BibitemShut {NoStop}%
\bibitem [{SM()}]{SM}%
  \BibitemOpen
  \href@noop {} {\emph {\bibinfo {title} {See Supplemental Material at
  http://link.aps.org/ supplemental/--/PhysRevLett., which includes details on
  the computational methods and on the conversion between computational and
  experimental time units, the caging time that a particle typically spend
  within a smectic layer as a function of its length, a discussion of the
  mechanism driving the adsorption to the smectic layer interfaces in the
  intermediate switching regime, and the potential barriers of the effective
  smectic potential as a function of $r$. The discussion includes Refs.
  \cite{Fre2001,Tir1979,Pat2012,Pat2015,Alv2017,Grel2014}.}}}\BibitemShut
  {Stop}%
\bibitem [{\citenamefont {van Bruggen}\ \emph {et~al.}(1998)\citenamefont {van
  Bruggen}, \citenamefont {Lekkerkerker}, \citenamefont {Maret},\ and\
  \citenamefont {Dhont}}]{Bru1998}%
  \BibitemOpen
  \bibfield  {author} {\bibinfo {author} {\bibfnamefont {M.~P.~B.}\
  \bibnamefont {van Bruggen}}, \bibinfo {author} {\bibfnamefont {H.~N.~W.}\
  \bibnamefont {Lekkerkerker}}, \bibinfo {author} {\bibfnamefont
  {G.}~\bibnamefont {Maret}}, \ and\ \bibinfo {author} {\bibfnamefont
  {J.~K.~G.}\ \bibnamefont {Dhont}},\ }\href {\doibase
  10.1103/PhysRevE.58.7668} {\bibfield  {journal} {\bibinfo  {journal} {Phys.
  Rev. E}\ }\textbf {\bibinfo {volume} {58}},\ \bibinfo {pages} {7668}
  (\bibinfo {year} {1998})}\BibitemShut {NoStop}%
\bibitem [{\citenamefont {L\"owen}(1999)}]{Low1999}%
  \BibitemOpen
  \bibfield  {author} {\bibinfo {author} {\bibfnamefont {H.}~\bibnamefont
  {L\"owen}},\ }\href {\doibase 10.1103/PhysRevE.59.1989} {\bibfield  {journal}
  {\bibinfo  {journal} {Phys. Rev. E}\ }\textbf {\bibinfo {volume} {59}},\
  \bibinfo {pages} {1989} (\bibinfo {year} {1999})}\BibitemShut {NoStop}%
\bibitem [{\citenamefont {Lettinga}\ \emph {et~al.}(2005)\citenamefont
  {Lettinga}, \citenamefont {Barry},\ and\ \citenamefont {Dogic}}]{Let2005}%
  \BibitemOpen
  \bibfield  {author} {\bibinfo {author} {\bibfnamefont {M.~P.}\ \bibnamefont
  {Lettinga}}, \bibinfo {author} {\bibfnamefont {E.}~\bibnamefont {Barry}}, \
  and\ \bibinfo {author} {\bibfnamefont {Z.}~\bibnamefont {Dogic}},\
  }\href@noop {} {\bibfield  {journal} {\bibinfo  {journal} {Europhysics
  Letters ({EPL})}\ }\textbf {\bibinfo {volume} {71}},\ \bibinfo {pages} {692}
  (\bibinfo {year} {2005})}\BibitemShut {NoStop}%
\bibitem [{\citenamefont {Alsayed}\ \emph {et~al.}(2005)\citenamefont
  {Alsayed}, \citenamefont {Islam}, \citenamefont {Zhang}, \citenamefont
  {Collings},\ and\ \citenamefont {Yodh}}]{Als2005}%
  \BibitemOpen
  \bibfield  {author} {\bibinfo {author} {\bibfnamefont {A.~M.}\ \bibnamefont
  {Alsayed}}, \bibinfo {author} {\bibfnamefont {M.~F.}\ \bibnamefont {Islam}},
  \bibinfo {author} {\bibfnamefont {J.}~\bibnamefont {Zhang}}, \bibinfo
  {author} {\bibfnamefont {P.~J.}\ \bibnamefont {Collings}}, \ and\ \bibinfo
  {author} {\bibfnamefont {A.~G.}\ \bibnamefont {Yodh}},\ }\href@noop {}
  {\bibfield  {journal} {\bibinfo  {journal} {Science}\ }\textbf {\bibinfo
  {volume} {309}},\ \bibinfo {pages} {1207} (\bibinfo {year}
  {2005})}\BibitemShut {NoStop}%
\bibitem [{\citenamefont {van~der Meer}\ \emph {et~al.}(2014)\citenamefont
  {van~der Meer}, \citenamefont {Qi}, \citenamefont {Fokkink}, \citenamefont
  {van~der Gucht}, \citenamefont {Dijkstra},\ and\ \citenamefont
  {Sprakel}}]{Meer2014}%
  \BibitemOpen
  \bibfield  {author} {\bibinfo {author} {\bibfnamefont {B.}~\bibnamefont
  {van~der Meer}}, \bibinfo {author} {\bibfnamefont {W.}~\bibnamefont {Qi}},
  \bibinfo {author} {\bibfnamefont {R.~G.}\ \bibnamefont {Fokkink}}, \bibinfo
  {author} {\bibfnamefont {J.}~\bibnamefont {van~der Gucht}}, \bibinfo {author}
  {\bibfnamefont {M.}~\bibnamefont {Dijkstra}}, \ and\ \bibinfo {author}
  {\bibfnamefont {J.}~\bibnamefont {Sprakel}},\ }\href {\doibase
  10.1073/pnas.1411215111} {\bibfield  {journal} {\bibinfo  {journal}
  {Proceedings of the National Academy of Sciences}\ }\textbf {\bibinfo
  {volume} {111}},\ \bibinfo {pages} {15356} (\bibinfo {year}
  {2014})}\BibitemShut {NoStop}%
\bibitem [{\citenamefont {van~der Meer}\ \emph {et~al.}(2017)\citenamefont
  {van~der Meer}, \citenamefont {Dijkstra},\ and\ \citenamefont
  {Filion}}]{Mee2017}%
  \BibitemOpen
  \bibfield  {author} {\bibinfo {author} {\bibfnamefont {B.}~\bibnamefont
  {van~der Meer}}, \bibinfo {author} {\bibfnamefont {M.}~\bibnamefont
  {Dijkstra}}, \ and\ \bibinfo {author} {\bibfnamefont {L.}~\bibnamefont
  {Filion}},\ }\href@noop {} {\bibfield  {journal} {\bibinfo  {journal} {The
  Journal of Chemical Physics}\ }\textbf {\bibinfo {volume} {146}},\ \bibinfo
  {pages} {244905} (\bibinfo {year} {2017})}\BibitemShut {NoStop}%
\bibitem [{\citenamefont {Belli}\ \emph {et~al.}(2010)\citenamefont {Belli},
  \citenamefont {Patti}, \citenamefont {van Roij},\ and\ \citenamefont
  {Dijkstra}}]{Bel2010}%
  \BibitemOpen
  \bibfield  {author} {\bibinfo {author} {\bibfnamefont {S.}~\bibnamefont
  {Belli}}, \bibinfo {author} {\bibfnamefont {A.}~\bibnamefont {Patti}},
  \bibinfo {author} {\bibfnamefont {R.}~\bibnamefont {van Roij}}, \ and\
  \bibinfo {author} {\bibfnamefont {M.}~\bibnamefont {Dijkstra}},\ }\href@noop
  {} {\bibfield  {journal} {\bibinfo  {journal} {The Journal of Chemical
  Physics}\ }\textbf {\bibinfo {volume} {133}},\ \bibinfo {pages} {154514}
  (\bibinfo {year} {2010})}\BibitemShut {NoStop}%
\bibitem [{\citenamefont {Naderi}\ \emph {et~al.}(2013)\citenamefont {Naderi},
  \citenamefont {Pouget}, \citenamefont {Ballesta}, \citenamefont {van~der
  Schoot}, \citenamefont {Lettinga},\ and\ \citenamefont {Grelet}}]{Nad2013}%
  \BibitemOpen
  \bibfield  {author} {\bibinfo {author} {\bibfnamefont {S.}~\bibnamefont
  {Naderi}}, \bibinfo {author} {\bibfnamefont {E.}~\bibnamefont {Pouget}},
  \bibinfo {author} {\bibfnamefont {P.}~\bibnamefont {Ballesta}}, \bibinfo
  {author} {\bibfnamefont {P.}~\bibnamefont {van~der Schoot}}, \bibinfo
  {author} {\bibfnamefont {M.~P.}\ \bibnamefont {Lettinga}}, \ and\ \bibinfo
  {author} {\bibfnamefont {E.}~\bibnamefont {Grelet}},\ }\href {\doibase
  10.1103/PhysRevLett.111.037801} {\bibfield  {journal} {\bibinfo  {journal}
  {Phys. Rev. Lett.}\ }\textbf {\bibinfo {volume} {111}},\ \bibinfo {pages}
  {037801} (\bibinfo {year} {2013})}\BibitemShut {NoStop}%
\bibitem [{\citenamefont {Lettinga}\ and\ \citenamefont
  {Grelet}(2007)}]{Let2007}%
  \BibitemOpen
  \bibfield  {author} {\bibinfo {author} {\bibfnamefont {M.~P.}\ \bibnamefont
  {Lettinga}}\ and\ \bibinfo {author} {\bibfnamefont {E.}~\bibnamefont
  {Grelet}},\ }\href {\doibase 10.1103/PhysRevLett.99.197802} {\bibfield
  {journal} {\bibinfo  {journal} {Phys. Rev. Lett.}\ }\textbf {\bibinfo
  {volume} {99}},\ \bibinfo {pages} {197802} (\bibinfo {year}
  {2007})}\BibitemShut {NoStop}%
\bibitem [{\citenamefont {Matena}\ \emph {et~al.}(2010)\citenamefont {Matena},
  \citenamefont {Dijkstra},\ and\ \citenamefont {Patti}}]{Mat2010}%
  \BibitemOpen
  \bibfield  {author} {\bibinfo {author} {\bibfnamefont {R.}~\bibnamefont
  {Matena}}, \bibinfo {author} {\bibfnamefont {M.}~\bibnamefont {Dijkstra}}, \
  and\ \bibinfo {author} {\bibfnamefont {A.}~\bibnamefont {Patti}},\ }\href
  {\doibase 10.1103/PhysRevE.81.021704} {\bibfield  {journal} {\bibinfo
  {journal} {Phys. Rev. E}\ }\textbf {\bibinfo {volume} {81}},\ \bibinfo
  {pages} {021704} (\bibinfo {year} {2010})}\BibitemShut {NoStop}%
\bibitem [{\citenamefont {Pouget}\ \emph {et~al.}(2011)\citenamefont {Pouget},
  \citenamefont {Grelet},\ and\ \citenamefont {Lettinga}}]{Pou2011}%
  \BibitemOpen
  \bibfield  {author} {\bibinfo {author} {\bibfnamefont {E.}~\bibnamefont
  {Pouget}}, \bibinfo {author} {\bibfnamefont {E.}~\bibnamefont {Grelet}}, \
  and\ \bibinfo {author} {\bibfnamefont {M.~P.}\ \bibnamefont {Lettinga}},\
  }\href {\doibase 10.1103/PhysRevE.84.041704} {\bibfield  {journal} {\bibinfo
  {journal} {Phys. Rev. E}\ }\textbf {\bibinfo {volume} {84}},\ \bibinfo
  {pages} {041704} (\bibinfo {year} {2011})}\BibitemShut {NoStop}%
\bibitem [{\citenamefont {Patti}\ \emph {et~al.}(2010)\citenamefont {Patti},
  \citenamefont {Masri}, \citenamefont {van Roij},\ and\ \citenamefont
  {Dijkstra}}]{Pat2010}%
  \BibitemOpen
  \bibfield  {author} {\bibinfo {author} {\bibfnamefont {A.}~\bibnamefont
  {Patti}}, \bibinfo {author} {\bibfnamefont {D.~E.}\ \bibnamefont {Masri}},
  \bibinfo {author} {\bibfnamefont {R.}~\bibnamefont {van Roij}}, \ and\
  \bibinfo {author} {\bibfnamefont {M.}~\bibnamefont {Dijkstra}},\ }\href@noop
  {} {\bibfield  {journal} {\bibinfo  {journal} {The Journal of Chemical
  Physics}\ }\textbf {\bibinfo {volume} {132}},\ \bibinfo {pages} {224907}
  (\bibinfo {year} {2010})}\BibitemShut {NoStop}%
\bibitem [{\citenamefont {Ruhwandl}\ and\ \citenamefont
  {Terentjev}(1996)}]{Ruh1996}%
  \BibitemOpen
  \bibfield  {author} {\bibinfo {author} {\bibfnamefont {R.~W.}\ \bibnamefont
  {Ruhwandl}}\ and\ \bibinfo {author} {\bibfnamefont {E.~M.}\ \bibnamefont
  {Terentjev}},\ }\href {\doibase 10.1103/PhysRevE.54.5204} {\bibfield
  {journal} {\bibinfo  {journal} {Phys. Rev. E}\ }\textbf {\bibinfo {volume}
  {54}},\ \bibinfo {pages} {5204} (\bibinfo {year} {1996})}\BibitemShut
  {NoStop}%
\bibitem [{\citenamefont {Stark}\ and\ \citenamefont
  {Ventzki}(2001)}]{Sta2001}%
  \BibitemOpen
  \bibfield  {author} {\bibinfo {author} {\bibfnamefont {H.}~\bibnamefont
  {Stark}}\ and\ \bibinfo {author} {\bibfnamefont {D.}~\bibnamefont
  {Ventzki}},\ }\href {\doibase 10.1103/PhysRevE.64.031711} {\bibfield
  {journal} {\bibinfo  {journal} {Phys. Rev. E}\ }\textbf {\bibinfo {volume}
  {64}},\ \bibinfo {pages} {031711} (\bibinfo {year} {2001})}\BibitemShut
  {NoStop}%
\bibitem [{\citenamefont {Loudet}\ \emph {et~al.}(2004)\citenamefont {Loudet},
  \citenamefont {Hanusse},\ and\ \citenamefont {Poulin}}]{Lou2004}%
  \BibitemOpen
  \bibfield  {author} {\bibinfo {author} {\bibfnamefont {J.~C.}\ \bibnamefont
  {Loudet}}, \bibinfo {author} {\bibfnamefont {P.}~\bibnamefont {Hanusse}}, \
  and\ \bibinfo {author} {\bibfnamefont {P.}~\bibnamefont {Poulin}},\
  }\href@noop {} {\bibfield  {journal} {\bibinfo  {journal} {Science}\ }\textbf
  {\bibinfo {volume} {306}},\ \bibinfo {pages} {1525} (\bibinfo {year}
  {2004})}\BibitemShut {NoStop}%
\bibitem [{\citenamefont {Kang}\ \emph {et~al.}(2005)\citenamefont {Kang},
  \citenamefont {Gapinski}, \citenamefont {Lettinga}, \citenamefont
  {Buitenhuis}, \citenamefont {Meier}, \citenamefont {Ratajczyk}, \citenamefont
  {Dhont},\ and\ \citenamefont {Patkowski}}]{Kan2005}%
  \BibitemOpen
  \bibfield  {author} {\bibinfo {author} {\bibfnamefont {K.}~\bibnamefont
  {Kang}}, \bibinfo {author} {\bibfnamefont {J.}~\bibnamefont {Gapinski}},
  \bibinfo {author} {\bibfnamefont {M.~P.}\ \bibnamefont {Lettinga}}, \bibinfo
  {author} {\bibfnamefont {J.}~\bibnamefont {Buitenhuis}}, \bibinfo {author}
  {\bibfnamefont {G.}~\bibnamefont {Meier}}, \bibinfo {author} {\bibfnamefont
  {M.}~\bibnamefont {Ratajczyk}}, \bibinfo {author} {\bibfnamefont {J.~K.~G.}\
  \bibnamefont {Dhont}}, \ and\ \bibinfo {author} {\bibfnamefont
  {A.}~\bibnamefont {Patkowski}},\ }\href {\doibase 10.1063/1.1834895}
  {\bibfield  {journal} {\bibinfo  {journal} {The Journal of Chemical Physics}\
  }\textbf {\bibinfo {volume} {122}},\ \bibinfo {pages} {044905} (\bibinfo
  {year} {2005})}\BibitemShut {NoStop}%
\bibitem [{\citenamefont {Mondiot}\ \emph {et~al.}(2012)\citenamefont
  {Mondiot}, \citenamefont {Loudet}, \citenamefont {Mondain-Monval},
  \citenamefont {Snabre}, \citenamefont {Vilquin},\ and\ \citenamefont
  {W\"urger}}]{Mon2012}%
  \BibitemOpen
  \bibfield  {author} {\bibinfo {author} {\bibfnamefont {F.}~\bibnamefont
  {Mondiot}}, \bibinfo {author} {\bibfnamefont {J.-C.}\ \bibnamefont {Loudet}},
  \bibinfo {author} {\bibfnamefont {O.}~\bibnamefont {Mondain-Monval}},
  \bibinfo {author} {\bibfnamefont {P.}~\bibnamefont {Snabre}}, \bibinfo
  {author} {\bibfnamefont {A.}~\bibnamefont {Vilquin}}, \ and\ \bibinfo
  {author} {\bibfnamefont {A.}~\bibnamefont {W\"urger}},\ }\href {\doibase
  10.1103/PhysRevE.86.010401} {\bibfield  {journal} {\bibinfo  {journal} {Phys.
  Rev. E}\ }\textbf {\bibinfo {volume} {86}},\ \bibinfo {pages} {010401}
  (\bibinfo {year} {2012})}\BibitemShut {NoStop}%
\bibitem [{\citenamefont {Turiv}\ \emph {et~al.}(2013)\citenamefont {Turiv},
  \citenamefont {Lazo}, \citenamefont {Brodin}, \citenamefont {Lev},
  \citenamefont {Reiffenrath}, \citenamefont {Nazarenko},\ and\ \citenamefont
  {Lavrentovich}}]{Tur2013}%
  \BibitemOpen
  \bibfield  {author} {\bibinfo {author} {\bibfnamefont {T.}~\bibnamefont
  {Turiv}}, \bibinfo {author} {\bibfnamefont {I.}~\bibnamefont {Lazo}},
  \bibinfo {author} {\bibfnamefont {A.}~\bibnamefont {Brodin}}, \bibinfo
  {author} {\bibfnamefont {B.~I.}\ \bibnamefont {Lev}}, \bibinfo {author}
  {\bibfnamefont {V.}~\bibnamefont {Reiffenrath}}, \bibinfo {author}
  {\bibfnamefont {V.~G.}\ \bibnamefont {Nazarenko}}, \ and\ \bibinfo {author}
  {\bibfnamefont {O.~D.}\ \bibnamefont {Lavrentovich}},\ }\href@noop {}
  {\bibfield  {journal} {\bibinfo  {journal} {Science}\ }\textbf {\bibinfo
  {volume} {342}},\ \bibinfo {pages} {1351} (\bibinfo {year}
  {2013})}\BibitemShut {NoStop}%
\bibitem [{\citenamefont {Martinez}\ \emph {et~al.}(2018)\citenamefont
  {Martinez}, \citenamefont {Collings},\ and\ \citenamefont {Yodh}}]{Mar2018}%
  \BibitemOpen
  \bibfield  {author} {\bibinfo {author} {\bibfnamefont {A.}~\bibnamefont
  {Martinez}}, \bibinfo {author} {\bibfnamefont {P.~J.}\ \bibnamefont
  {Collings}}, \ and\ \bibinfo {author} {\bibfnamefont {A.~G.}\ \bibnamefont
  {Yodh}},\ }\href {\doibase 10.1103/PhysRevLett.121.177801} {\bibfield
  {journal} {\bibinfo  {journal} {Phys. Rev. Lett.}\ }\textbf {\bibinfo
  {volume} {121}},\ \bibinfo {pages} {177801} (\bibinfo {year}
  {2018})}\BibitemShut {NoStop}%
\bibitem [{\citenamefont {Grelet}(2014)}]{Grel2014}%
  \BibitemOpen
  \bibfield  {author} {\bibinfo {author} {\bibfnamefont {E.}~\bibnamefont
  {Grelet}},\ }\href {\doibase 10.1103/PhysRevX.4.021053} {\bibfield  {journal}
  {\bibinfo  {journal} {Phys. Rev. X}\ }\textbf {\bibinfo {volume} {4}},\
  \bibinfo {pages} {021053} (\bibinfo {year} {2014})}\BibitemShut {NoStop}%
\bibitem [{\citenamefont {Bolhuis}\ and\ \citenamefont
  {Frenkel}(1997)}]{Bol1997}%
  \BibitemOpen
  \bibfield  {author} {\bibinfo {author} {\bibfnamefont {P.}~\bibnamefont
  {Bolhuis}}\ and\ \bibinfo {author} {\bibfnamefont {D.}~\bibnamefont
  {Frenkel}},\ }\href {\doibase 10.1063/1.473404} {\bibfield  {journal}
  {\bibinfo  {journal} {The Journal of Chemical Physics}\ }\textbf {\bibinfo
  {volume} {106}},\ \bibinfo {pages} {666} (\bibinfo {year}
  {1997})}\BibitemShut {NoStop}%
\bibitem [{\citenamefont {Grelet}(2008)}]{Gre2008}%
  \BibitemOpen
  \bibfield  {author} {\bibinfo {author} {\bibfnamefont {E.}~\bibnamefont
  {Grelet}},\ }\href {\doibase 10.1103/PhysRevLett.100.168301} {\bibfield
  {journal} {\bibinfo  {journal} {Phys. Rev. Lett.}\ }\textbf {\bibinfo
  {volume} {100}},\ \bibinfo {pages} {168301} (\bibinfo {year}
  {2008})}\BibitemShut {NoStop}%
\bibitem [{\citenamefont {Patti}\ and\ \citenamefont {Cuetos}(2012)}]{Pat2012}%
  \BibitemOpen
  \bibfield  {author} {\bibinfo {author} {\bibfnamefont {A.}~\bibnamefont
  {Patti}}\ and\ \bibinfo {author} {\bibfnamefont {A.}~\bibnamefont {Cuetos}},\
  }\href {\doibase 10.1103/PhysRevE.86.011403} {\bibfield  {journal} {\bibinfo
  {journal} {Phys. Rev. E}\ }\textbf {\bibinfo {volume} {86}},\ \bibinfo
  {pages} {011403} (\bibinfo {year} {2012})}\BibitemShut {NoStop}%
\bibitem [{\citenamefont {Cuetos}\ and\ \citenamefont {Patti}(2015)}]{Pat2015}%
  \BibitemOpen
  \bibfield  {author} {\bibinfo {author} {\bibfnamefont {A.}~\bibnamefont
  {Cuetos}}\ and\ \bibinfo {author} {\bibfnamefont {A.}~\bibnamefont {Patti}},\
  }\href {\doibase 10.1103/PhysRevE.92.022302} {\bibfield  {journal} {\bibinfo
  {journal} {Phys. Rev. E}\ }\textbf {\bibinfo {volume} {92}},\ \bibinfo
  {pages} {022302} (\bibinfo {year} {2015})}\BibitemShut {NoStop}%
\bibitem [{\citenamefont {van Roij}\ \emph {et~al.}(1995)\citenamefont {van
  Roij}, \citenamefont {Bolhuis}, \citenamefont {Mulder},\ and\ \citenamefont
  {Frenkel}}]{Roi1995}%
  \BibitemOpen
  \bibfield  {author} {\bibinfo {author} {\bibfnamefont {R.}~\bibnamefont {van
  Roij}}, \bibinfo {author} {\bibfnamefont {P.}~\bibnamefont {Bolhuis}},
  \bibinfo {author} {\bibfnamefont {B.}~\bibnamefont {Mulder}}, \ and\ \bibinfo
  {author} {\bibfnamefont {D.}~\bibnamefont {Frenkel}},\ }\href {\doibase
  10.1103/PhysRevE.52.R1277} {\bibfield  {journal} {\bibinfo  {journal} {Phys.
  Rev. E}\ }\textbf {\bibinfo {volume} {52}},\ \bibinfo {pages} {R1277}
  (\bibinfo {year} {1995})}\BibitemShut {NoStop}%
\bibitem [{\citenamefont {van Duijneveldt}\ and\ \citenamefont
  {Allen}(1997)}]{Dui1997}%
  \BibitemOpen
  \bibfield  {author} {\bibinfo {author} {\bibfnamefont {J.~S.}\ \bibnamefont
  {van Duijneveldt}}\ and\ \bibinfo {author} {\bibfnamefont {M.~P.}\
  \bibnamefont {Allen}},\ }\href@noop {} {\bibfield  {journal} {\bibinfo
  {journal} {Molecular Physics}\ }\textbf {\bibinfo {volume} {90}},\ \bibinfo
  {pages} {243} (\bibinfo {year} {1997})}\BibitemShut {NoStop}%
\bibitem [{\citenamefont {Siavashpouri}\ \emph {et~al.}(2019)\citenamefont
  {Siavashpouri}, \citenamefont {Sharma}, \citenamefont {Fung}, \citenamefont
  {Hagan},\ and\ \citenamefont {Dogic}}]{Sia2019}%
  \BibitemOpen
  \bibfield  {author} {\bibinfo {author} {\bibfnamefont {M.}~\bibnamefont
  {Siavashpouri}}, \bibinfo {author} {\bibfnamefont {P.}~\bibnamefont
  {Sharma}}, \bibinfo {author} {\bibfnamefont {J.}~\bibnamefont {Fung}},
  \bibinfo {author} {\bibfnamefont {M.~F.}\ \bibnamefont {Hagan}}, \ and\
  \bibinfo {author} {\bibfnamefont {Z.}~\bibnamefont {Dogic}},\ }\href@noop {}
  {\bibfield  {journal} {\bibinfo  {journal} {Soft Matter}\ }\textbf {\bibinfo
  {volume} {15}},\ \bibinfo {pages} {7033} (\bibinfo {year}
  {2019})}\BibitemShut {NoStop}%
\bibitem [{\citenamefont {Frenkel}\ and\ \citenamefont {Smit}(2001)}]{Fre2001}%
  \BibitemOpen
  \bibfield  {author} {\bibinfo {author} {\bibfnamefont {D.}~\bibnamefont
  {Frenkel}}\ and\ \bibinfo {author} {\bibfnamefont {B.}~\bibnamefont {Smit}},\
  }\href@noop {} {\emph {\bibinfo {title} {Understanding molecular simulation:
  from algorithms to applications}}}\ (\bibinfo  {publisher} {Elsevier},\
  \bibinfo {year} {2001})\BibitemShut {NoStop}%
\bibitem [{\citenamefont {Tirado}\ and\ \citenamefont {de~la
  Torre}(1979)}]{Tir1979}%
  \BibitemOpen
  \bibfield  {author} {\bibinfo {author} {\bibfnamefont {M.~M.}\ \bibnamefont
  {Tirado}}\ and\ \bibinfo {author} {\bibfnamefont {J.~G.}\ \bibnamefont {de~la
  Torre}},\ }\href {\doibase 10.1063/1.438613} {\bibfield  {journal} {\bibinfo
  {journal} {The Journal of Chemical Physics}\ }\textbf {\bibinfo {volume}
  {71}},\ \bibinfo {pages} {2581} (\bibinfo {year} {1979})}\BibitemShut
  {NoStop}%
\end{thebibliography}%

\end{document}